# Shorter duration of slow wave sleep is related to symptoms of depression in patients with epilepsy

Stefanía Eyjolfsdóttir[1]

Eugen Trinka[2]

Yvonne Höller[1]

[1]Faculty of Psychology, University of Akureyri, Akureyri, Iceland;
[2]Department of Neurology, Christian Doppler University Hospital, Paracelsus Medical University and Centre for Cognitive Neuroscience Salzburg, Austria. Member of the European Reference Network EpiCARE. Neuroscience Institute, Christian Doppler University Hospital, Paracelsus Medical University and Centre for Cognitive Neuroscience Salzburg, Austria.

Correspondence: Yvonne Höller

ORCID 0000-0002-1727-8557

Norðurslóð 2, 600 Akureyri, Iceland

Tel +354 460 8576

Email : yvonne@unak.is

**Abstract:**
Slow wave sleep duration and spectral abnormalities are related to both epilepsy and depression, but it is unclear how depressive symptoms in patients with epilepsy are affected by slow wave sleep duration and clinical factors, and how the spectral characteristics of slow wave sleep reflect a potential interaction of epilepsy and depression.

Long-term video-EEG monitoring was conducted in 51 patients with focal epilepsy, 13 patients with generalized epilepsy, and 9 patients without epilepsy. Slow wave sleep segments were manually marked in the EEG and duration as well as EEG power spectra were extracted. Depressive symptoms were documented with the Beck Depression Inventory (BDI).

At least mild depressive symptoms (BDI>9) were found among 23 patients with focal epilepsy, 5 patients with generalised epilepsy, and 6 patients who had no epilepsy diagnosis. Slow wave sleep duration was shorter for patients with at least mild depressive symptoms (p=.004), independently from epilepsy diagnosis, antiseizure medication, age, and sex. Psychoactive medication was associated with longer slow wave sleep duration (p=.008). Frontal sigma band power (13-15 Hz) during slow wave sleep was higher for patients without epilepsy and without depressive symptoms as compared to patients without depressive symptoms but with focal epilepsy (p=.005).

Depressive symptoms affect slow wave sleep duration of patients with epilepsy similarly as in patients without epilepsy. Since reduced slow wave sleep can increase the likelihood of seizure occurrence, these results stress the importance of adequate treatment for patients with epilepsy who experience depressive symptoms.

**Keywords:** epilepsy, slow wave sleep, depression, EEG sigma band

# 1   Introduction

Around 50 million people of all ages, genders, and social classes worldwide are affected by epilepsies.[1,2] Epilepsies are a heterogeneous group of disorders and are one of the most common neurological disorders worldwide, often associated with comorbid psychiatric disorders.[3,4] The prevalence of epilepsy is higher in low/middle income countries compared to high/middle income countries and high-income countries. [2,5,6]

Epilepsy and sleep have a complex relationship, influenced by sleep stage, epilepsy syndrome, and antiseizure medication (ASM).[7] People with epilepsy experience sleep disturbances more often than healthy controls. Specifically, epilepsy was shown to affect sleep efficiency in terms of more sleep stage shifts, higher number and longer duration of awakenings.[8,9] Epilepsy patients have shorter total sleep time, decreased sleep efficiency, longer latency to reach rapid eye movement (REM) sleep, and less time spent in REM sleep as compared to healthy controls.[10] Excessive daytime sleepiness, insomnia, sleep disordered breathing, such as obstructive sleep apnea, restless legs syndrome, narcolepsy, and insufficient sleep have been reported to affect people with epilepsy.[11,12] Furthermore, ASM impacts sleep, e.g. by increasing (gabapentin, lamotrigine, perampanel, pregabalin and tiagabine) or reducing (phenobarbitone) the duration of slow wave sleep.[13] Additionally, research on sleep oscillations in patients with epilepsy needs to take ASM into account as it affects frequency content of brain activity.[14]

Sleep problems among patients with epilepsy deserve special attention since lack of adequate sleep and sleep deprivation is well known to trigger seizures.[7] Thus, difficulty in achieving adequate sleep can result in increased frequency of seizure events that in return disturb sleep patterns, ending in a vicious cycle.[15] However, the balance between waking and sleeping states can either increase or decrease the tendency of seizures and interictal discharges.[11] Furthermore, when considering individual sleep stages it turns out that non-REM is known to facilitate both seizure onset and seizure spread while REM sleep is known for inhibiting epileptic activity.[7,16]

Epileptic activity was also shown to have a direct impact on sleep architecture and sleep oscillations.[16] Slow wave sleep activity as an index of physiologic sleep can be measured as spectral power in the delta band, which is enhanced in patients with epilepsy depending on the lateralization of the epileptic focus.[10,17,18] Specifically, affected regions in patients with focal epilepsy show an increase of slow wave activity both during wakefulness and sleep in comparison with healthy controls.[17] Patients with focal epilepsy also show locally unspecific higher power in the alpha, sigma, and beta band during sleep.[17] We can conclude that epileptic processes interact with the expression of sleep markers,[18] and alterations of sleep are particularly evident for non-REM sleep e.g. in the form of non-REM sleep instability.[19]

It was previously suggested that impaired sleep of patients with epilepsy is rather caused by the co-morbidity of depression but not by epilepsy itself.[20] Indeed, epilepsy is often co-morbid with depression[21, 23] which has well documented effects on sleep.[24] It was previously reported that patients with epilepsy show a correlation between poor sleep quality and elevated symptoms of depression.[25] Depression can increase the frequency of seizures by inducing sleep deprivation.[26] Sleep architecture is affected by depression, especially in the form of a reduced duration of slow wave sleep.[27,28] Reduced slow wave sleep worsens mood in patients with major depressive disorder.[29] This result was interpreted as an indicator that there may be an impairment in the homeostatic mechanism that modulates synaptic strength in people with major depressive disorder, and the severity of this impairment can be associated with the severity of mood disturbances.[29] In the interplay of depression and sleep it is also important to consider that it was reported previously that depressed men show less slow-wave sleep than depressed women.[30] Most interestingly, during slow wave sleep, patients with major depression show reduced amplitude of delta activity[31] and individuals at risk show increased sigma activity[32]. However, while these aspects are documented for patients with depression, it is unclear how epilepsy and depression interact with respect to slow wave sleep duration and frequency content of slow wave sleep. A potential association of slow wave sleep duration and depression in patients with epilepsy should be clarified to help determining the causes of depression in epilepsy, with the aim of providing more targeted treatment. Also, the knowledge of epilepsy and depression as contributing factors on specific frequency alterations in slow wave sleep would contribute to a better understanding of the factors that affect poor sleep of patients with epilepsy. Therefore, we aimed to answer the following research questions:

Is slow wave sleep duration related to depressive symptoms in patients with epilepsy?

Are power spectra of slow wave sleep in patients with epilepsy differentially affected by depressive symptoms?

## 2 Material and methods

### 2.1 Ethics

The project obtained prior approval from the ethical commission of the Region of Salzburg, Austria (approval nr. 415-E/2044/35-2019). All participants gave written informed consent with the opportunity to discuss the research with both a doctor and a researcher. The Ethical principles according to the Declaration of Helsinki were followed throughout the project.

## 2.2 Setting

The epilepsy monitoring unit consists of four beds in one room for simultaneous observation of four patients through video and EEG recordings, monitored for 24 hours each day by medical staff. Admitted patients undergo diagnostic evaluations that consist of long-term video EEG monitoring from Monday afternoon through Friday morning. Patients are admitted to the epilepsy monitoring unit to classify the epilepsy syndrome, for differential diagnosis of suspicious events, to assess seizure frequency, to optimize medication, or for presurgical evaluation.

Daily rhythm was standardized for all patients, with specific time ranges that consists of lights on (06:30-07:00), breakfast (07:00-07:30), lunch (11.30), dinner (16:30), and lights off (22:00-00:00). During the monitoring period it is common to taper the dosage of antiepileptic drugs and expose patients to sleep deprivation in the third night to provoke a timely occurrence of seizures.

## 2.3 Sample recruitment

This study was based on electroencephalography (EEG) data from patients who sought treatment at the Epilepsy Monitoring Unit at the Department of Neurology, Christian Doppler Medical Center, Salzburg, Austria. Other data obtained from this sample was analysed and published previously.[33,34]

Inclusion criteria were i) age >=18 years, ii) diagnosis of epilepsy according to best clinical evidence (i.e. converging clinical evidence, interictal 24h EEG and magnetic resonance imaging or single-photon emission computerized tomography).

Exclusion criteria were previous neurosurgical treatment of seizures as these patients also come to the epilepsy monitoring unit for a follow-up. Among the four patients admitted per week, we chose weekly the best suited patient according to these exclusion and inclusion criteria, and also by avoiding patients about whom it is already known that they most likely experience psychogenic seizures. There were 106 patients that were able to give written informed consent and were admitted to the epilepsy monitoring unit between the 2nd of February 2016 and the 11th of June 2018 were considered for recruitment.

From this sample, patients were excluded if ii) they experience status epilepticus during the stay, ii) no slow-wave sleep segments found in the second or third night of the stay, iii) medically requested sleep deprivation in the assessed night.

## 2.4 Depression screening

Depression screening was performed by a psychologist with the second revision of the German translation of the Beck Depression Inventory (BDI).[35] The questions were presented on a computer screen and the patients were asked to respond via button press. The BDI is a 21-item self-report scale that measures depressive symptoms and severity in people of age 13 years and older.[36] It rates severity of depression symptoms on a 4-point scale from 0 (symptom absent) to 3 (severe symptom).[37] Recall period for items is the last 2 weeks to correspond with Diagnostic and Statistical Manual of Mental Disorders, Fourth Edition (DSM – IV).[38] Scoring is achieved by summing up the highest rating for each item on the scale, the minimum score is 0 and the maximum score is 63.[37] We used a cutoff of 10 which is considered to indicate at least mild depression.[39] This threshold was found to demonstrate good sensitivity and specificity with medical inpatients with a sensitivity of 83% and specificity of 65%[40] and is therefore a good screening cutoff for depression[41].

## 2.5 EEG Data acquisition

Routine procedure in the epilepsy monitoring unit consists of 24 hours recording of video-EEG using a Micromed S.p.A (https://micromedgroup.com/) system, called SystemPlus Evolution and with SD LTM 64 Express Amplifier. Twenty-nine electrodes were used and placed by trained medical staff according to the international 10-20 system, that involves Fpz that served as ground, and Oz as reference. Additional electrodes measured horizontal electrooculogram, chin-electromyogram, and electrocardiogram from the chest. Impedances were kept below 10 kΩ. Sampling rate was 1,024-Hz, filtered by 0,1-Hz high pass and 50-Hz notch.

## 2.6 Identification of slow wave sleep

Night files from the epilepsy monitoring unit were extracted and concatenated and imported into the software Brain Vision Analyzer (Brain Products GmbH), which allows scanning trough EEG recordings, with 20 seconds epochs displayed at the screen, and manually placing markers that indicate start and end of slow wave sleep segments in each patient's recordings, alongside with further options for quantitative EEG analysis. An epoch length of 20-30 seconds was originally recommended [42], and today's standard is 30 seconds [43], but shorter epochs are recommendable for research purposes when sleep is fragmented [44], which is typical in epilepsy patients [16]. Slow wave sleep, corresponding to stages 3 and 4 in the Rechtschaffen and Kales manual [42] and to stage N3 in the new AASM scoring manual [43], was assessed visually based on all available channels of the frequency below 2 Hz and defined as incidence of low-frequency oscillations in the EEG with high amplitude, synchronized activity.[45, 47] Moreover, slow waves must occupy more than 50% of an epoch to be judged as a slow wave sleep.[47] However, magnitude and frequency of the slow wave sleep EEG signals depends on the individual that is being measured.[48] Night 2 of each patient was chosen for analysis, in order to discard night one which is often biased by adjustment of the patient to the recording environment.[49] If night 2 resulted in no deep sleep, or if the patient underwent sleep deprivation in night 2, night 3 was reviewed. If the same conditions applied to night 3, the patient was excluded. All EEG segments were reviewed and marked by a trained researcher, and then reviewed by an experienced EEG specialist who adjusted markers, if necessary.

## 2.7 EEG analysis

From the resulting deep sleep segments, total duration of deep sleep was extracted. Power spectral analysis was conducted as follows: Deep sleep segments were segmented into sub-segments of 4sec. Each sub-segment was transformed to frequency domain using the Fast Fourier Transform (resolution 0.25 Hz, non-complex output in microvoltage, half spectrum). Then, the resulting power spectra were averaged over all available sub-segments. From these power spectra averages power density was extracted as unsigned, rectified values and mean activity for the frequency bands delta (0.5-4 Hz), theta (5-7 Hz), alpha (8-12 Hz), sigma (13-15 Hz), beta (16-29 Hz), and gamma (30-45 Hz). To consider the spatial distribution of EEG power, values were averaged for the lobes and hemispheres, resulting in the areas frontal left (F3, F7, F11), frontal right (F4, F8, F12), central left (C3), central right (C4), temporal left (T7, T11), temporal right (T8, T12), parietal left (P3, P7, P11), parietal right (P4, P8, P12), occipital left (O1), and occipital right (O2).

## 2.8   *Antiseizure medication and psychoactive medication.*

Low-dose combinations of multiple drugs were reported to be better tolerated than a high dose of a single drug.[50, 52] Therefore, we calculated the drug load by hands of the defined daily dose for epilepsy medication and psychoactive medication as we described in a previously published study.[34] Drug load was calculated for the day of admission as well as the day before the investigated night. First, we calculated the ratio between the prescribed dose and the defined daily dose separately for ASM and psychoactive medication. Then, we summed those loads over all drugs taken on that day, again separately for ASM and psychoactive drugs. Other medications such as for treatment of hypertension, hormonal birth control etc. were not considered.

## 2.9   *Statistics*

Statistics were done with R/R-Studio.[53] We used logistic regression with the R function *glm* to find significant determinants of depression. Depression was a binary outcome variable, split according to the BDI threshold 10 (<10: no depression; >=10 at least mild depression). Included determining factors were slow wave sleep duration, epilepsy diagnosis (focal, generalized, no epilepsy), duration of epilepsy (difference between age at testing and age of onset), localization of the epileptic focus (frontal, temporal, other), sex, as well as drug load of epileptic and psychoactive drugs on the day before the analyzed night. We added Spearman correlation coefficients for the correlations between the ordinal variables and BDI score to ease interpretation of the results.

For statistical analysis of EEG power spectra, the following repeated-measures factors were used: hemisphere (left, right), frequency (delta, theta, alpha, sigma, beta, and gamma), and lobe (frontal, temporal, central, parietal, occipital). The same grouping variable for depression was used again according to the BDI threshold of 10, indicating at least mild depression. To test for differences of depression groups in the frequency spectrum of slow wave sleep we conducted a semi-parametric repeated measures ANOVA that allows for non-normality and variance heterogeneity[54] using the function *RM* from the package MANOVA.RM.[55] Please note that since this method allows for non-normal error terms and heteroscedastic variance, we did not additionally need to normalize the power of the individual frequency bands by the sum of the spectrum. We report Wald-Type Statistics alongside with the parametric resampling p-value obtained after 1000 iterations, since the sample sizes are very unequally distributed. Significant main effects and interactions that were relevant for our research questions were followed up with post-hoc tests that were conducted as univariate ANOVA with the same method. Bonferroni-Holm correction was applied in all instances of multiple comparisons.[56]

# 3  Results

## 3.1  Sample

From initially 106 participants in the study, seven patients were excluded because of incomplete EEG data and five patients were excluded because no deep sleep was found, neither in night 2 nor in night 3. Furthermore, 11 patients had to be excluded because they did not undergo depression screening with the BDI. Finally, ten patients were excluded because their epilepsy diagnosis remained unclear after the five-day epilepsy video-EEG monitoring. In the final sample of N=73 patients there were 36 women (mean age = 32.28; SD = 11.96) and 37 men (mean age 32; SD = 14.5). For 66 patients, night two was assessed. In two patients, night 2 yielded no deep sleep segments, such that night 3 was assessed. Furthermore, five patients underwent sleep deprivation during night 2, and we therefore chose to assess night 3, instead. There were eight patients who were left-handed, six of them with epilepsy.

After the 5-day video-EEG monitoring, diagnoses were focal epilepsy (N=51), generalized epilepsy (N=13), and 9 were diagnosed to not have epilepsy. Diagnoses in this sample were panic attacks (2), convulsive syncope (1), transient disorder of consciousness of unknown etiology (1), migraine (2), cardiovascular disorder (1), pneumonia (1), and spinal fracture (1). Among the epilepsy patients with focal epilepsy, localization was temporal left in 13 cases, 10 cases had temporal right localization, two temporal bilateral, 14 frontal, two occipital, three parietal, and six had an unclear localization. Among the patients with epilepsy, onset was on average at the age of 21.9 years (SD=14.37) and ranged from 3 months to 62 years. Among patients with focal epilepsy, 18 showed structural abnormalities according to magnetic resonance neuroimaging (see supplementary table for details) while only one among the patients with generalized epilepsy showed structural abnormalities (1cm sized arachnoid cyst at the left insular cistern, not related to the epilepsy syndrome). Furthermore, three patients with focal epilepsy underwent brain surgery in the past (1 parahippocampal arterio-venous malformation, 2) neuroepithelial tumor, 3) ganglioglioma right occipitotemporal gyrus). On admission, 13 epilepsy patients took no medication, 32 were on ASM monotherapy, 16 had a combination of two ASMs and five took three ASMs. No patient took barbiturates. Among all patients, five were on psychoactive medication with only one type of medication, one took two types, and three took three types of psychoactive medication. Overall, the psychoactive drugs taken were antidepressants (all 9 patients), antipsychotics (4 patients), benzodiazepines (3 patients), anxiolytics (1 patients), and opioids (1 patient). Drug load of ASM for epilepsy patients was on average 1.28 (SD=1.29) on admission, but because of drug tapering during epilepsy monitoring, this average decreased to 0.73 (SD=1.04) on the day before the investigated night. A full list of medications taken, and dosage is provided in the supplementary file.

## 3.2 Depression

For patients with epilepsy the median of the BDI score was 7 (range 0-39) while for patients without epilepsy the median of the BDI score was 12 (range 2-42). Elevated BDI scores (BDI score of at least 10) were found in 23 patients with focal epilepsy, five patients with generalized epilepsy, and six patients who had no epilepsy diagnosis.

## 3.3 Duration of deep sleep

Average duration of slow wave sleep segments was 77.84 min (SD=43.53) for patients with epilepsy and 63.04 (SD=23.26) for patients without epilepsy. Patients with low levels of depressive symptoms spent on average 86.73 min (SD=45.84) in slow wave sleep while patients with elevated depressive symptoms spent on average 63.72 min (SD=33.02) in slow wave sleep. The logistic regression model which determined depression by slow wave sleep duration, drug load of ASM and psychoactive drugs, age of onset, age, sex, localization of the epileptic focus and epilepsy diagnosis with 5 Fisher Scoring iterations had an Akaike Information Criterion of 87.37. Coefficients are shown in Table 1.

**Table 1:** Logistic regression coefficients and Spearman correlation coefficients (rho) for correlations between factor and BDI score.

| Factors | Estimate | SE | z-value | p-value | rho |
|---|---|---|---|---|---|
| Slow wave sleep duration | -0.04 | 0.01 | -2.86 | .004 | -.180 |
| Frontal vs. other localization | -0.18 | 0.88 | -0.20 | .843 | |
| Frontal vs. temporal localization | -0.40 | 0.84 | -0.48 | .633 | |
| Duration of epilepsy | 0.12 | 0.05 | 2.25 | .025 | -.190 |
| Antiseizure medication drug load | -0.11 | 0.32 | -0.36 | .717 | -.220 |
| Psychoactive drug load | 2.44 | 0.92 | 2.65 | .008 | .430 |
| Sex | -0.56 | 0.64 | -0.87 | .383 | |

| | | | | |
|---|---|---|---|---|
| Focal epilepsy | 0.09 | 1.24 | 0.08 | .940 |
| Generalized epilepsy | -1.37 | 1.45 | -0.95 | .344 |

**Abbreviations:** SE, Standard Error;

According to the results as shown in Table 1, elevated depression scores were significantly related to shorter duration of slow wave sleep, and this relation was independent from ASM drug load, sex, focus localization, and epilepsy diagnosis. However, a higher psychoactive drug load and a later age of onset was associated significantly with elevated depression scores. It should be noted that longer duration of epilepsy was related significantly to reduced depressive symptomatology.

We additionally tested for effects of elevated depressive symptoms and epilepsy diagnosis on the frequency spectrum of slow wave sleep. Table 2 shows the results.

**Table 2:** Results of the semi-parametric ANOVA with the grouping factors depression (low, high) and epilepsy (focal, generalized, no epilepsy), and the repeated measures factors lobe (frontal, central, temporal, parietal, occipital), frequency (delta, theta, alpha, sigma, beta, and gamma), and hemisphere (left, right). Wald-Type statistical p-values are shown alongside with the parametric resampling p-value

| Effect/interaction | F | df | p-value | BS-p |
|---|---|---|---|---|
| depression | 0.43 | 1 | .511 | .521 |
| epilepsy | 3.58 | 2 | .167 | .294 |
| depression x epilepsy | 2.80 | 2 | .247 | .382 |
| lobe | 664.74 | 4 | <.001 | <.001 |
| depression x lobe | 4.73 | 4 | .317 | .424 |
| epilepsy x lobe | 13.03 | 8 | .111 | .418 |
| depression x epilepsy x lobe | 10.05 | 8 | .261 | .539 |
| frequency | 1445.42 | 5 | <.001 | <.001 |
| depression x frequency | 22.09 | 5 | .001 | .073 |

| | F | df | p | BS-p |
|---|---|---|---|---|
| epilepsy x frequency | 21.57 | 10 | .017 | .376 |
| depression x epilepsy x frequency | 47.96 | 10 | <.001 | .089 |
| lobe x frequency | 2894.62 | 20 | <.001 | <.001 |
| depression x lobe x frequency | 100.00 | 20 | <.001 | .081 |
| epilepsy x lobe x frequency | 971.10 | 40 | <.001 | .096 |
| depression x epilepsy x lobe x frequency | 1235.18 | 40 | <.001 | .064 |
| hemisphere | 0.01 | 1 | .924 | .929 |
| depression x hemisphere | 0.03 | 1 | .853 | .857 |
| epilepsy x hemisphere | 1.79 | 2 | .41 | .451 |
| depression x epilepsy x hemisphere | 1.48 | 2 | .478 | .503 |
| lobe x hemisphere | 4.36 | 4 | .359 | .475 |
| depression x lobe x hemisphere | 3.00 | 4 | .558 | .627 |
| epilepsy x lobe x hemisphere | 17.11 | 8 | .029 | .218 |
| depression x epilepsy x lobe x hemisphere | 5.10 | 8 | .747 | .879 |
| frequency x hemisphere | 11.52 | 5 | .042 | .180 |
| depression x frequency x hemisphere | 5.61 | 5 | .346 | .510 |
| epilepsy x frequency x hemisphere | 4.72 | 10 | .909 | .964 |
| depression x epilepsy x frequency x hemisphere | 14.76 | 10 | .140 | .550 |
| lobe x frequency x hemisphere | 454.07 | 20 | <.001 | <.001 |
| depression x lobe x frequency x hemisphere | 33.76 | 20 | .028 | .692 |
| epilepsy x lobe x frequency x hemisphere | 835.62 | 40 | <.001 | .085 |
| depression x epilepsy x lobe x frequency x hemisphere | 664.39 | 40 | <.001 | .100 |

**Abbreviations:** F, test value; df, degrees of freedom; BS-p, bootstrapped resampling p-value;

The classical Wald-Type statistic yielded significant group interactions, but none of the effects that included at least one of the grouping factors depression or epilepsy were significant after obtaining a p-value with resampling, considering the unequal sample sizes. However, several group effects and interactions showed a tendency for significance at p<.100 even in the resampling p-value. The interactions between 1) depression and frequency, 2) depression, epilepsy, and frequency, 3) depression, lobe, and frequency, 4) epilepsy, lobe, and frequency, 5) depression, epilepsy, lobe, and frequency, and 6) epilepsy, lobe, frequency, and hemisphere. Since we were interested in an interaction between epilepsy and depression, we performed post-hoc tests for the interactions 2) depression, epilepsy, and frequency, as well as 5) depression, epilepsy, lobe, and frequency. While no significant differences were found in the delta, theta, alpha, beta, and gamma band, the sigma band was the only band that showed a significant p-value according to the resampling post-hoc test method when testing group interactions for each frequency band (depression x epilepsy: $F(20)=272.36$; $p<.001$; resampling $p= .039$) where the classical Wald-Type statistic was significant after correcting for multiple comparisons, but the resampled p-value was not significant after correcting for multiple comparisons. When further distinguishing by lobe, the only effects that had a significant p-value according to the resampling method were again found in the sigma band, and here in the frontal lobe (depression x epilepsy: $F(4)=26.51$; $p<.001$; resampling $p=.021$) and the parietal lobe (epilepsy: $F(4)=20.05$; $p<.001$; resampling $p= .046$). Again, these post-hoc tests were significant only according to the classical Wald-Type statistic but not according to the resampled p-value. The interaction for depression x epilepsy in the frontal lobe and sigma band is illustrated in Figure 1.

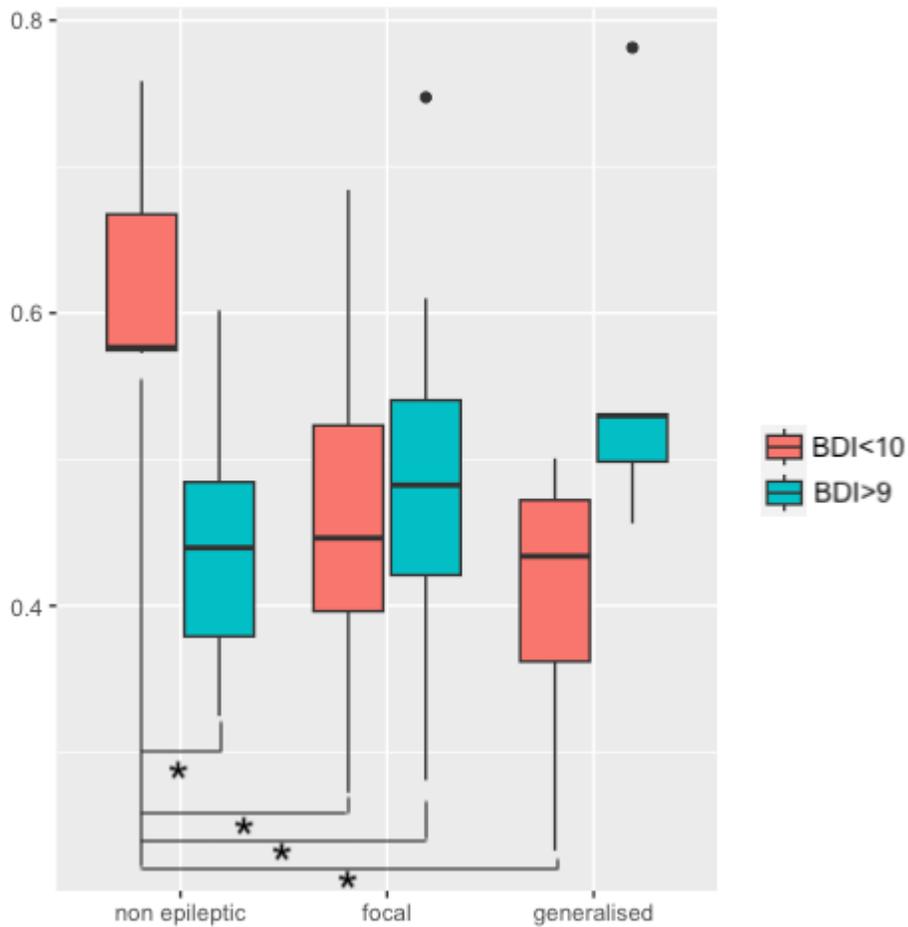

**Figure 1**: Boxplots of power in the sigma band during slow wave sleep derived from frontal EEG channels (F3, F7, F11, F4, F8, and F12) for the significant interaction between groups by depression and epilepsy diagnosis. Boxes indicate the interquartile range, the horizontal bold line the median. Whiskers indicate the range from the minimum to the maximum value, except for the case of outliers, which are defined by extending beyond the interquartile range by more than 1.5 times the interquartile range.

Post-hoc tests revealed a higher sigma band power in the frontal lobe for patients without epilepsy and without depressive symptoms as compared to all other groups except for the group with at least mild depression (BDI>9) and generalized epilepsy. This difference was significant after correcting for multiple comparisons for the comparison to patients without depression and with focal epilepsy (see Table 3).

**Table 3:** Significant Wilcoxon post-hoc tests for frontal sigma power

| No depression, no epilepsy vs. ... | W | p-value |
|---|---|---|
| …depression, no epilepsy | 135.5 | .016 |
| …no depression, focal epilepsy | 80 | **.005** |
| …no depression, generalized epilepsy | 24 | .012 |
| …depression, focal epilepsy | 63 | .036 |

**Abbreviations:** W, Wilcoxon's W as test value;

**Note:** Bold font indicates significance after correction for multiple comparisons

# 4 Discussion

In this study we found that elevated depression scores were related to shorter duration of slow wave sleep for all patients, i.e., this effect was similar for patients with focal or generalized epilepsy, as well as no epilepsy diagnosis. This effect was independent from ASM drug load and sex, longer epilepsy duration was related to elevated depressive symptoms. Also, higher psychoactive drug load was associated with higher depressive symptoms. Additionally, we found that patients without depressive symptoms and without epilepsy showed higher frontal power in the sigma band.

According to our results, the relation between slow wave sleep duration and depression is not specific to patients with epilepsy, or, in other words, epilepsy patients are not more vulnerable to develop depressive symptoms linked to short slow wave sleep duration than people without epilepsy. Although the relation between slow wave sleep and depression was independent of epilepsy, this does not mean that sleep problems can be disregarded in patients with epilepsy. Our results support previous claims that depression and sleep disturbance among patients with epilepsy deserve clinical attention,[20] especially in view of a high rate of suicidal ideation among patients with epilepsy that is associated poor sleep,[57] the high prevalence of depression,[4,22,23,58] and the impact on quality of life.[59] There is even indication that not only is epilepsy a risk factor for depression, but depression is also a risk factor for more severe epilepsy. [60]

Age and sex are factors that interact with both depression and sleep.[30,61,62] More specifically, increased age is a risk factor for depression in patients with epilepsy.[63] While sex did not play a role for depressive symptoms according to our data, we found that longer duration of epilepsy was linked to lower depression scores. While this seems counterintuitive at the first sight, an additional analysis (Supplementary File 2) showed that both age at onset as well as by tendency also age at the time of testing are correlated with higher depression scores. Thus, the relationship with duration might be skewed by sample characteristics. Furthermore, we found no effect of ASM on depressive symptoms, but there was an association between a higher psychoactive drug load and elevated levels of depressive symptoms, which is most plausibly explained by the fact that antidepressants were the most commonly taken medication, i.e., all 9 patients who took psychoactive medication took antidepressants. It is also well documented that antidepressants affect sleep[64,65] In previous research it was found that epilepsy patients with moderate depressive symptoms are more likely to use two or more types of ASM, and that specifically topiramate or clonazepam were more commonly used among depressed patients[63]. We did not use medication count but determined drug load, which is a more detailed approach to assess the effect of medication but might certainly lead to different results. Furthermore, we did not investigate the impact of individual types of medication as this would have led to rather small subgroups. Specifically, clonazepam was used by only one patient and topiramate was used only in two cases in our sample, of which one had low depressive symptoms and the other one had elevated depressive symptoms.

With respect to our findings from the spectral analysis of slow wave sleep, it is interesting to note that our findings in the sigma band point to a possible effect on sleep spindles since the sigma power frequency band includes background EEG activity as well as the power of sleep spindles.[66] Sleep spindles are sensitive to barbiturates and sedative drugs and interact with epilepsy type (generalized vs. focal).[67] Specifically, in generalized epilepsy the thalamo-cortical system is abnormally excited with an increased likelihood of spindles to occur, [68] with an increased likelihood over the frontal cortex for some forms of generalized epilepsy. [69] In patients with focal epilepsy, sleep spindle rates are globally and focally lower as compared to controls, [70] which might explain our results. Furthermore, sleep spindles are enhanced in patients with depression. [71,72] Sleep spindles and slow wave activity are inversely interrelated, such that the power of sleep spindles is higher in sleep stage N2 than N3 [66]. This suggests that the effect we found might be even clearer if we would have included stage N2 into the analysis.

In addition to the differences in sleep spindle activity between depressed and non-depressed individuals, sigma power during non-REM sleep was previously reported to be elevated in non-depressed individuals who have a positive family history of depression as compared to patients with major depression and as compared to healthy controls without risk factors for depression[73] suggesting that it might indicate a general vulnerability for depression. In contrast, our results suggest elevated sigma power in participants without depression and without epilepsy. However, neither did we include a clinical diagnosis of major depression but only elevated depression scores, nor did we control for family history of depression, which might explain the diverging result. Nevertheless, both the present data as well as prior research[73] suggest the relevance of sigma band power during slow wave sleep as a correlate of central nervous system changes in relation to depression. The reactivity of frontal sigma is of particular interest as the prefrontal cortex was suggested to be the part of the brain that relies on slow wave sleep as a stage of cerebral recovery, which might be dysfunctional in affective disorders such as depression.[61]

Frontal differences between epilepsy patients with and without depression in the gamma range were reported recently but using the Kullback-Leibler divergence in contrast to the here reported spectral power analysis.[74]

Frontal lobe functional and structural abnormalities are a frequent finding in patients with major depressive disorder, but also among those at risk.[75] The vulnerability is likely linked to frontally located dysfunctions, such as negative processing biases and learned helplessness[75] which are important targets for psychotherapeutic intervention. It was also found that depressed patients show a smaller decrease in frontal metabolism from pre-sleep wakefulness to NREM sleep [76]. This finding was interpreted as nonrestorative sleep in the relevant networks [76]. We could speculate that frontal lobe abnormality as found in people with epilepsy and depressive symptoms during deep sleep reflects dysfunctional cognitive processes being caused by inadequate recovery of the relevant brain circuits – a hypothesis that is worth further investigation. However, our results did not indicate a particular vulnerability of patients with frontal lobe epilepsy.

Despite asymmetric power distributions were previously reported to differ between controls and depressed patients during NREM sleep [77] we did not find an effect of hemisphere as included in our EEG analysis model. This is likely explained by the circumstance of the large number of patients with focal epilepsy which is left- or right- lateralized, adding additional variance to the data. We did not consider lateralization of epileptic focus as an additional factor in the model because this would have resulted in a rather small sample size for the subgroups.

### *4.1 Limitations*

A main limitation of the present study is the small sample size of the subgroups with uneven distribution, which resulted from convenience sampling as patients were enrolled consecutively in the epilepsy monitoring unit. However, we took this aspect into account by performing additional statistical evaluation with parametric bootstrap resampling that reveals more conservative and more accurate p-values. Still, the small sample sizes prevented us from performing more specific subgroup analyses. For example, it would have been desirable to examine subgroups based on the type of medications taken, but this approach must be taken in future studies with larger sample sizes.

Although all participants in the study underwent the same procedure, the sleep conditions might have varied to some extent. Specifically, because patients sleep in the epilepsy monitoring unit in a room with 4 beds, the presence of other patients with sleep-associated events can disturb naturally occurring sleep patterns in the included patients. Also, analysis of the night after the sleep-deprivation night might impact deep sleep duration.

A further limitation can be construed from the way we screened for depression instead of performing a diagnostic interview. It was reported previously that general depression screenings might not be suited for patients with epilepsy and that measures that are specifically designed for medically ill patients would be a better choice.[78] In relation to this limitation, the number of patients with clinically relevant depression was rather low. This limits the comparability of our data to previous studies which investigated patients with major depression.

While we chose to identify slow-wave sleep segments manually, automated sleep scoring will most likely replace this time-consuming task in the future.[79] Automated sleep scoring might be challenging in epilepsy, as epileptiform activity can hamper the applicability of algorithms that have been developed for healthy sleep staging. Specifically, some epileptic activity during sleep can resemble normal variants that might be indicative for certain sleep stages.[80]

Another limitation of our analysis is that we did not account for changes in spectral power over the night and, thus, across the sleep cycles, but extracted the overall frequency spectrum from all deep sleep segments together. Analysis of power across the night may lead to a better understanding of sleep homeostasis in patients with epilepsy and the interplay with depressive symptoms.

Finally, while we accounted for multiple confounding factors, the influence of comorbid medical conditions and mental disorders, medication refractoriness, and seizure frequency were not considered.

# 5 Conclusion

It is difficult to determine the exact interrelationship between sleep, epilepsy, and depression, since they all are risk factors for each other. We might speculate that these three conditions suffer from impairments in the same networks, an idea that could be investigated in future research. However, the importance of adequate sleep cannot be overestimated for patients with epilepsy, especially with respect to the high prevalence of depression among patients with epilepsy that is strongly connected to poor sleep.[58]

According to our data, shorter duration of slow wave sleep is associated to elevated depressive symptoms among patients with epilepsy, and longer duration of epilepsy is additionally contributing to depression. Although patients with epilepsy do not differ from patients without epilepsy in respect to slow wave sleep duration, our findings well support the claim that patients with epilepsy should get adequate treatment of sleep problems in order to reduce their vulnerability for depressive symptoms.

# 6 Acknowledgments

Thanks to the students Miriam Bacher, Nina Biller, Björn Butter, Lukas Christmann, Leonie Dehne, Isabelle Ehrlich, Maike Engel, Sophia Gabler, Johannes Göller, Daniel Göller, Dirk Gütlin, Lilian Gutsch, Viola Heberger, Lucie Knörr, Pascal Lempe, Johanna Luxbauer, Adrian Marcu, Valentina Ostreljanovic, Lucas Rainer, Christoph Schabernag, Frank van Schalkwijk, Denise Schindlmaier, Ronny Sluka, Anja Ströhlein, Marina Thierauf, Tamara Traub, Elisabeth Weingartner, and Philipp Windhager for help with the intensive examination schedule. Thanks to Christina Florea, Bernhard Ganser, Constantin Hecker, Magdolna Mezes, Tobias Moser, Zsolt Nagy, Caroline Neuray, Ferdinand Otto, Rudolf Kreidenhuber, Alexander Kunz, David Gabelia, Alexandra Rohracher, Fabio Rossini, Markus Leitinger, Judith Dobesberger, Julia Höfler, and Gudrun Kalss for help with recruitment and providing informed consent. Thanks to Eydís Anna Guðmundsdóttir for pre-reviewing a couple of night recordings. Thanks to Michaela Pötzelsberger and the staff at the epilepsy monitoring unit for help, support, and patience with our daily visits for examination, the weekly recruitment, and for sharing their office space with the scientific staff.

# 7 Data availability

Because of the sensitive nature of the data, further information beyond what is provided in the supplementary section it is made available upon reasonable request and, if necessary, after approval of the relevant ethical authority.

## 8 Conflicts of interest

ET reports personal fees from EVER Pharma, Marinus, Arvelle, Angelini, Argenx, Medtronic, Bial-Portela & Cª, New Bridge, SK, Pharma, GL Pharma, GlaxoSmithKline, Boehringer Ingelheim, LivaNova, Eisai, UCB, Biogen, Sanofi, Jazz Pharma, and Actavis. His institution received grants from Biogen, UCB Pharma, Eisai, Red Bull, Merck, Bayer, the European Research Council, Austrian Research Fund (FWF), Bundesministerium für Wissenschaft und Forschung, and Jubiläumsfond der Österreichischen Nationalbank.

YH reports fees for lectures from Eisai. She received prior funding from the Icelandic Research Fund, Austrian Research Fund, and the European Research Council.

SE reports no conflicts of interest.

## 9 Funding

This work was supported by the Austrian Science Fund (FWF): T 798-B27 and by the Research Fund of the Paracelsus Medical University (PMU-FFF): A-16/02/021-HÖL.

## 10 Supplementary Material

The supplementary file 1 lists all included patients, including information which night was analyzed, age at testing, gender, handedness, BDI score, epilepsy type, and medication on day of admission. The supplementary table 2 represents an additional analysis of the logistic regression, separating duration of epilepsy into the factors age at testing and age at onset.